\documentclass[%
 reprint,
 amsmath,amssymb,
 aps,
]{revtex4-2}

\usepackage{graphicx}
\usepackage{dcolumn}
\usepackage{bm}
\usepackage[utf8]{inputenc}
\usepackage{ulem}

\usepackage{amsmath}
\usepackage{hyperref}
\usepackage{cleveref}
\usepackage{xparse}
\usepackage{physics}
\usepackage[caption=false]{subfig}
\usepackage{graphicx}
\usepackage{xcolor}

\usepackage{csquotes}

\newcommand{\lucio}[1]{\textcolor{black}{#1}}

\begin{document}

\preprint{APS/123-QED}

\title{Dynamic Phase Alignment in Navier-Stokes Turbulence}

\author{Lucio M. Milanese}
\email{milanese@mit.edu}
\affiliation{Plasma Science and Fusion Center, Massachusetts Institute of Technology, Cambridge, MA 02139, USA}

\author{Nuno F. Loureiro}
\affiliation{Plasma Science and Fusion Center, Massachusetts Institute of Technology, Cambridge, MA 02139, USA}

\author{Stanislav Boldyrev}
\affiliation{Department of Physics, University of Wisconsin-Madison, Madison, WI 53706, USA}
\affiliation{Space Science Institute, Boulder, CO 80301, USA}

\date{\today}

\begin{abstract}

In Navier-Stokes turbulence, energy and helicity injected at large scales are subject to a joint direct cascade, with both quantities exhibiting a spectral scaling $\propto k^{-5/3}$. We demonstrate via direct numerical simulations that the two cascades are compatible due to the existence of a strong scale-dependent phase alignment between velocity and vorticity fluctuations, with the phase alignment angle scaling as $\cos\alpha_k\propto k^{-1}$.
\end{abstract}

\maketitle

\textit{Introduction.}
The incompressible Navier-Stokes equations (NSEs) govern the dynamics of a broad variety of physical systems \cite{Foias2001Navier-StokesTurbulence}. Many of these systems are in a turbulent state: they exhibit chaotic dynamics which cannot be readily described from first principles, but is instead partially captured by phenomenological and statistical models \cite{Frisch1995Turbulence}. The prominent Kolmogorov cascade model \cite{Kolmogorov1941TheNumber} captures key aspects of fluid turbulent dynamics based on the assumption that\lucio{, in three-dimensional systems,} the kinetic energy of the turbulent fluctuations is transferred from larger to smaller-scale structures via nonlinear interactions that are local in wavenumber $k$.

Kolmogorov's theory of turbulence was developed considering energy as the only nonlinear invariant of the system. It was subsequently discovered \cite{Moreau1961ConstantesBarotrope, Moffatt1969TheLines}, however, that a second inviscid invariant of the incompressible NSEs exists, namely, the helicity, defined as $\mathcal{H} = \int \vb{v} \cdot  \vb{\boldsymbol{\omega}} \ dV,
\label{helicity}$ with $\vb{v}$ the velocity of the fluid and $\vb{\boldsymbol{\omega}} = \nabla \times \vb{v}$ its vorticity. A flow with net helicity is necessarily more complex than otherwise, as its mirror symmetry is broken~\cite{Chen2003IntermittencyHelicityc}.

The existence of a second nonlinear invariant further complicates the analysis of the turbulent dynamics.~In principle, it is possible for both invariants to cascade forward, i.e., from large to small scales, or for them to cascade in different directions \cite{Alexakis2018CascadesFlowsb, Pouquet2019HelicityReviewb}. When both quantities cascade forward, it is in principle possible for either invariant to set the (scale-dependent) amplitude of velocity fluctuations, thus affecting the nonlinear eddy turnover time and leading to different predictions \cite{Brissaud1973HelicityTurbulence, Alexakis2018CascadesFlowsb}. Indeed, the helicity density at scale $\lambda$ can be dimensionally evaluated as  $H_\lambda\sim v_\lambda \omega_\lambda \sim v_\lambda^2/\lambda$, and the nonlinear interaction time (eddy turnover time) as $\tau_\lambda\sim \lambda/v_\lambda$, where $v_{\lambda}$ and $\omega_{\lambda}$ denote the typical amplitudes of velocity and vorticity fluctuations at scale $\lambda$. Assuming the existence of a constant helicity flux, i.e., $\varepsilon_H\sim H_\lambda/\tau_\lambda\sim {\rm const.}$, one would derive $v_\lambda\propto \lambda^{2/3}$, and the resulting energy and helicity spectra would be, respectively, $\mathcal{E}(k) \propto k^{-7/3}$ and $\mathcal{H}(k)\propto k^{-4/3}$ \cite{Brissaud1973HelicityTurbulence}. This scenario leads to a scaling of the energy spectrum different from Kolmogorov's  $\mathcal{E}(k) \propto k^{-5/3}$, which is instead defined by a constant energy flux, $\varepsilon\sim v_\lambda^2/\tau_\lambda$, and the corresponding velocity scaling $v_\lambda\propto \lambda^{1/3}$. In this case, dimensional arguments would predict the spectral scaling of helicity to be $\mathcal{H}(k) \propto k^{-2/3}$. However, this ``na\"ive" estimate of the helicity spectrum is, as we discuss below, inconsistent with a constant helicity flux in the inertial range. 

In many instances, the simultaneous presence of two invariants in a turbulent system requires that one conserved quantity cascades to small scales, while the other one cascades to large scales.~Such a phenomenon was discovered by Kraichnan in pioneering work on two-dimensional turbulence \cite{Kraichnan1967InertialTurbulence}, and later studied more broadly in various models of weak and strong turbulence \cite{Frisch1975PossibilityTurbulence,Hasegawa1985Self-organizationMedia,Zakharov1992KolmogorovTurbulence}.~In the case of Navier-Stokes turbulence, the na\"ive dimensional arguments suggest that it is the helicity invariant that should exhibit the direct cascade, while the energy should inverse cascade.~Indeed, a cascade of energy to small scales, $\varepsilon = \textnormal{const.}$, would seem to imply the divergence of the helicity flux, $\varepsilon_H\sim \varepsilon/\lambda$ at small scales, contradicting helicity conservation.~Similarly, if helicity cascaded to large scales, $\varepsilon_H=\textnormal{const.}$, then the energy flux $\varepsilon\sim \lambda \varepsilon_H$ would diverge at large scales, contradicting energy conservation. The only possibility of maintaining a steady state would be, therefore, to assume a direct cascade for helicity and inverse for energy. 

This is not, however, what occurs in three-dimensional Navier-Stokes turbulence, where instead \textit{both} energy and helicity are observed to cascade forward.~Theoretical arguments in favor of the existence of a joint direct cascade of the two invariants in the presence of net helicity were first put forward in Ref.~\cite{Brissaud1973HelicityTurbulence}, based on conservation of energy and helicity in the inertial range, and in Ref.~\cite{Kraichnan1973HelicalEquilibrium}, based on the analysis of inviscid statistical equilibria.~There exists today significant experimental and numerical evidence that energy and helicity in Navier-Stokes turbulence are indeed both subject to a direct cascade \cite{Borue1997SpectraTurbulence,Chen2003IntermittencyHelicityc,Koprov2005ExperimentalLayer,Alexakis2018CascadesFlowsb,Qu2018CascadesTurbulence,Pouquet2019HelicityReview, Plunian2020InverseTurbulenceb}, and share a spectral scaling of $\propto k^{-5/3}$.  While the double direct cascade of energy and helicity has been the subject of robust investigation and seems beyond reasonable doubt \cite{Chen2003TheTurbulence,Chen2003IntermittencyHelicityc,Eyink2005LocalityCascades,Baerenzung2008SpectralHelicity,Stepanov2009SpectralTurbulence,Choi2009AlignmentTurbulence,Teitelbaum2009EffectFlows,Plunian2011HelicityScalings,Pouquet2013InverseFlows,Biferale2013SplitTurbulence,Gledzer2015InverseModes,DePietro2015InverseTurbulence,Sahoo2015RoleFluctuations,Stepanov2015HinderedTurbulence,Imazio2017PassiveTurbulence,Sahoo2017HelicityModels,Chkhetiani2017HelicalGeneration,Briard2017ClosureTurbulence,Teimurazov2018DirectCode,Sahoo2018EnergyEquations,Yan2020ScaleSpace,Plunian2020InverseTurbulenceb}, the turbulent mechanism that enables it has remained unclear.

The goal of this Letter is to uncover such mechanism.~We argue that this joint cascade is possible because the velocity and vorticity fluctuations develop a progressively stronger \textit{phase correlation} at smaller scales.
More precisely, we propose that while the energy flux is dimensionally evaluated as $\varepsilon\sim  v^2_{\lambda}/ \tau_{\lambda}$, 
the helicity flux should be evaluated as
\lucio{\begin{eqnarray}
\varepsilon_H\sim r_{\lambda} v_{\lambda} \omega_{\lambda} /\tau_{\lambda} \sim r_{\lambda} v^2_{\lambda} /(\lambda \tau_{\lambda}),
\end{eqnarray}}
where $r_{\lambda}$ is a scale-dependent cancellation factor.~Assuming that the (scale-dependent) nonlinear time $\tau_\lambda$ is determined purely by the energy cascade, and requiring that both the energy and helicity fluxes be constant in the inertial range, allows one to predict the scaling of the cancellation factor:
\begin{equation}
r_{\lambda} \sim \lambda/L ,
\end{equation}
where $L$ is the outer scale of the turbulence.

When the cancellation factor is present, the simultaneous direct cascade of \textit{both} energy and helicity becomes possible, and one predicts the scaling $\mathcal{H}(k)dk \propto k^{-5 / 3} d k$ for the helicity spectrum.~\lucio{In this Letter we argue that the cancellation factor underpinning the spectral scaling of helicity is a manifestation of \textit{dynamic phase alignment}, i.e., a scale-dependent correlation between the fluctuations of velocity and vorticity, and demonstrate this result by means of direct numerical simulations of driven, incompressible Navier-Stokes turbulence.}

\textit{Numerical setup}. We integrate the NSEs numerically with the pseudospectral code \texttt{Tarang} \cite{Verma2013BenchmarkingSimulations, Chatterjee2018ScalingCoresb} on a cubic, triply periodic domain using a grid of $N^3$ collocation points. \lucio{The ``2/3's rule" is used for dealiasing \cite{Verma2013BenchmarkingSimulations}.}~The model equations read 
\begin{equation}
    \frac{\partial \mathbf{v}}{\partial t}+\mathbf{v} \cdot {\bm\nabla} \mathbf{v}=-{\bm\nabla} P+ \nu {\nabla}^{2} \mathbf{v}+\mathbf{F}, \label{eq:nse_one}
\end{equation}
coupled to the incompressibility condition, ${\bm \nabla} \cdot \mathbf{v} = 0$.~$P$ is the pressure, $\nu$ is the kinematic viscosity and $\mathbf{F}$ represents external forcing.~Both $P$ and $\mathbf{F}$ are normalized to the fluid density, set as $\rho\equiv 1$. The pressure is computed by solving a Poisson equation obtained by taking the divergence of Eq.~(\ref{eq:nse_one}) and using the incompressibility condition.~In all simulations, energy and helicity are injected at wavenumbers $2 \leq|\mathbf{k}_f| \leq 6$. Wavenumbers are normalized to the size of the simulation domain, so that the smallest wavenumber, which represents box-scale fluctuations, has value unity. We define the energy and helicity injection rates, respectively, as
\begin{gather}
    \epsilon_{E}=\sum_{\mathbf{k}_f} \mathfrak{R}\left\{\mathbf{F}(\mathbf{k}) \cdot \mathbf{v}^{*}(\mathbf{k})\right\}, \\
   \epsilon_{H}=\frac{1}{2} \sum_{\mathbf{k}_f} \mathfrak{R}\left\{\mathbf{F}(\mathbf{k}) \cdot \boldsymbol{\omega}^{*}(\mathbf{k})-\mathbf{v}(\mathbf{k}) \cdot(i \mathbf{k} \times \mathbf{F}(\mathbf{k}))\right\},
\end{gather}
where $\mathbf{F}(\mathbf{k})$ represents the delta-correlated in time forcing term, and $\mathfrak{R}$ denotes the real part.~The forcing algorithm is described in detail in Ref.~\cite{Teimurazov2018DirectCode}.~We further define the ratio of helicity to energy injection as $\mathcal{R}_{\mathcal{H}} = \epsilon_H/\bar{k} \epsilon_{E}$, where $\bar{k} = \sum_{\mathbf{k}_f} |\mathbf{k}|F(\mathbf{k})/\sum_{\mathbf{k}_f}F(\mathbf{k})$ is the weight-averaged wavenumber of the forcing. 

Table \ref{tab:simulation_parameters} summarizes key parameters of the simulations performed.~We define the Reynolds number as $\textnormal{Re} = v_{rms}L/\nu$, where  $L = \int_{0}^{\infty} k^{-1} \mathcal{E}(k) d k / \int_{0}^{\infty} \mathcal{E}(k) d k$ is the integral length scale of the turbulence and $v_{rms} = \sqrt{2\mathcal{E}/3}$ is the root mean square of the velocity fluctuations. The inverse of the Kolmogorov scale is represented by $k_d = (\nu^3/\epsilon_E)^{-1/4}$.  
\begin{table}[h]
\begin{ruledtabular}
\begin{tabular}{ccccc}
ID & N & Re & $\mathcal{R}_{\mathcal{H}}$ & $k_d$ \\
\colrule
A1 & 1024 & 1350 & 0.1 & 150 \\
A2 & 1024 & 1350 & 0.3 & 150 \\ 
A3 & 1536 & 2000 & 0.5 & 205 \\
\end{tabular}
\caption{\label{tab:simulation_parameters}%
Summary of key simulation parameters.}
\end{ruledtabular}
\end{table}

\begin{figure*}
\centering 
\includegraphics[width=0.99\textwidth]{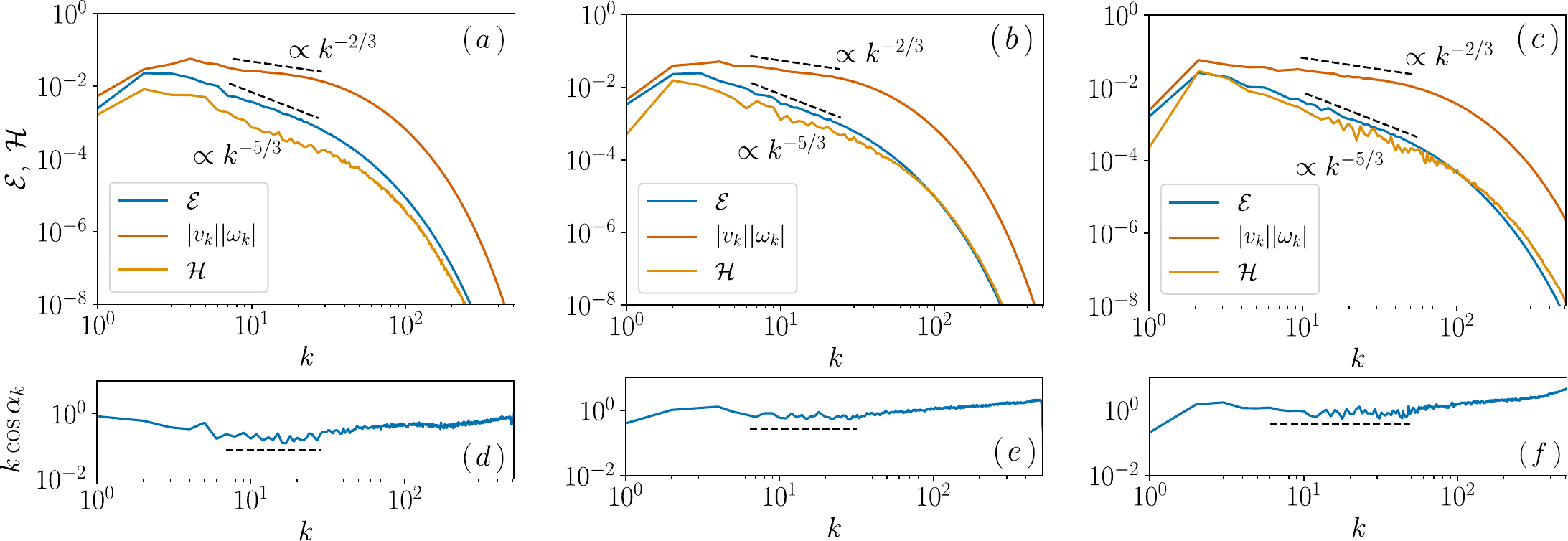}
\caption{Subplots (a,b,c) show spectra of energy, helicity and the product of the absolute value of velocity and vorticity for simulations A1-A3, respectively. Subplots (d,e,f) present \lucio{compensated spectra of} the average value of the cosine of the phase angle (Eq.~(\ref{eq:cos_alpha})) between velocity and vorticity fluctuations. The inertial range covers wavenumbers up to $k \approx 35$ (for simulations A1 and A2) and up to $k \approx 50$ (for A3), after which dissipation becomes non-negligible and the predicted scalings do not apply.}
\label{fig:spectra_and_alignment}
\end{figure*}

\textit{Dynamic phase alignment.} The net helicity density at each wavenumber is a function of the absolute value of the Fourier coefficients $|v_{k}|$ and $|\omega_{k}|$, and of the phase angle between them, $\alpha_k$. At a given wavenumber $k$, the average value of $\alpha_k$ is given by
\begin{equation}
\label{eq:cos_alpha}
\cos\alpha_k = \frac{1}{2} \sum_{i} \frac{\langle v_{{i}_{k}} \omega_{{i}_{k}}^{*} + c. c. \rangle}{\langle|v_{{i}_{k}}||\omega_{{i}_{k}}|\rangle},
\end{equation}
where $\langle ... \rangle$ represents averaging over the wavenumber shell and we are summing over the index $i \in \{x, y, z\}$ representing the three spatial directions.~\lucio{We can write the spectral scaling of helicity as $\mathcal{H}(k) \sim k^{-1} v_{\lambda} \omega_{\lambda} \cos \alpha_k \sim k^{-2/3} \cos \alpha_k$, where the last step is obtained under the assumptions $v_{\lambda} \sim k^{-1/3}$ and $\omega_{\lambda} \sim k^{2/3}$.~Conservation of helicity in the direct cascade requires a scaling $\mathcal{H}(k) \sim k^{-5/3}$. We thus predict a scaling $\cos \alpha_k \sim k^{-1}$.}

In Fig.~\ref{fig:spectra_and_alignment}(a-c), we show the energy and helicity spectra (from time-averaged data in steady state) for simulations A1-A3. Good agreement with the scaling $\propto k^{-5/3}$ is recovered for both invariants. The product of the absolute value of velocity and vorticity (the denominator of the right-hand side in Eq.~\ref{eq:cos_alpha}) exhibits good agreement with the spectral scaling $\propto k^{-2/3}$, predicted via dimensional arguments.~Fig. \ref{fig:spectra_and_alignment}(d-f) shows \lucio{compensated spectra of} the phase alignment angle between fluctuations of velocity and vorticity as function of scale, demonstrating that it follows a scaling $\cos \alpha_k \propto k^{-1}$ in the inertial range. The plots in Fig.~\ref{fig:spectra_and_alignment} demonstrate that the scaling $\cos \alpha_k \propto k^{-1}$ occurs for a broad range of ratios of helicity to energy injection in the system. \lucio{An investigation of the Reynolds number dependence of these results is reported in the Supplemental Material.}

When $\cos\alpha_k=0$, helicity is zero and the system is mirror symmetric. The scaling $\cos\alpha_k \propto k^{-1}$ therefore implies that dynamic phase alignment underpins the restoration of mirror symmetry at small scales, in agreement with simulations of Navier-Stokes turbulence where net helicity is injected at large scales \cite{Chen2003IntermittencyHelicityc,Alexakis2018CascadesFlowsb,Pouquet2019HelicityReviewb}. 

\textit{Absence of geometric angle alignment.}~We argued above that the reduction of helicity was associated with phase alignment, that is, scale-dependent cancellation of positive and negative values of the product $\mathbf{v} \cdot \bm{\omega}$.~\lucio{However, a scale-dependent \textit{angular} alignment between the directions of the fields may also, in principle, constitute the \textit{primary} mechanism of helicity reduction.}~In order to investigate the possible existence of such an angular alignment, we consider the band-pass filtered velocity and vorticity fields.~\lucio{The band-pass filtered velocity is defined as $\mathbf{v}_{\ell}=\mathcal{F}^{-1}(\mathbf{\hat{v}})$, where $\mathcal{F}^{-1}$ denotes the inverse Fourier transform, while $\mathbf{\hat{v}}$ denotes the field after undergoing Gaussian filtering of the form $\exp{-(k-k_\textnormal{c})^2/2\sigma^2}$, with  $k_{\textnormal{c}}^{-1} \equiv  \ell$ and $\sigma = 3$.} The band-pass filtered vorticity field is defined as $\boldsymbol{\omega}_{\ell}= \nabla \times \mathbf{v}_{\ell}$ \footnote{We use Fourier band-pass filtering to determine $\mathbf{v}_{\ell}$ and $\boldsymbol{\omega}_{\ell}$ because, as is well-known \cite{Ott1992Sign-singularTurbulence, Vainshtein1994Sign-singularScalings,Sahoo2017HelicityModels}, two-point structure functions of vorticity are heavily influenced by small-scale fluctuations and thus cannot be reliably used to compare different scales in the inertial range.}.~We will demonstrate numerically that their phase and angular correlations disentangle statistically, and that, in contrast with phase alignment, geometric angular alignment does not exhibit scale dependence.
Consider first two scalar fields, $X$ and $Y$, that may stand for, e.g., the x-components of the velocity and vorticity vectors. Assume that we may represent them as $X=X_0\cos\phi$ and $Y=Y_0\cos(\phi+\phi_0)$, where $X_0$ and $Y_0$ are positive amplitudes, $\phi$ is a random phase, and $\phi_0$ is the phase shift.~The correlation between these fields is given by $\langle XY\rangle=(1/2)X_0Y_0\cos\phi_0$. Similarly, one can calculate the correlation between the normalized fields, $X/|X|$ and $Y/|Y|$. The product of these fields can only take two values, $+1$ or $-1$, with the corresponding probabilities depending on the phase shift, $P_+=1-(\phi_0/\pi)$ and $P_-=\phi_0/\pi$. It is convenient to denote the deviation of the phase shift from $\pi/2$ as ${\tilde \phi_0}=\pi/2-\phi_0$, so that the probabilities take the form $P_+=1/2+{\tilde \phi}_0/\pi$ and $P_-=1/2-{\tilde \phi}_0/\pi$. Then, the correlation function of the normalized fields is 
\begin{equation}
\Big\langle\frac{X}{|X|}\frac{Y}{|Y|} \Big\rangle=P_+-P_-=\frac{2{\tilde \phi_0}}{\pi}.
\end{equation}

We now turn to the vector fields $\mathbf{v}_{\ell}$ and $\boldsymbol{\omega}_{\ell}$. In order to study their phase and geometric correlations, consider the product  $Z=\mathbf{v}_{\ell}\cdot\boldsymbol{\omega}_{\ell}/|\mathbf{v}_{\ell}||\boldsymbol{\omega}_{\ell}|$. Clearly, we can represent it as $Z=\xi |\cos\theta|$, where the ``phase" variable $\xi$ takes the values $\pm 1$, while $\theta$ is the geometric angle between the directions of the fields. The statistics of $Z$ are given by the joint probability density function $p(\xi, |\cos\theta|)$:
\begin{eqnarray}
p\left(\xi, |\cos\theta|\right)=p\left(\xi\Big| |\cos\theta|\right)p\left(|\cos\theta|\right).
\end{eqnarray}
The probability density of the variable $Z$ is then:
\begin{eqnarray}\label{ProbaZ}
p(Z)=\left\{
\begin{matrix}
& p\left(+1\Big||\cos\theta|\right)p\left(|\cos\theta|\right), \quad Z>0,\\
& p\left(-1\Big||\cos\theta|\right)p\left(|\cos\theta|\right), \quad Z<0,
\end{matrix}
\right.\quad\quad
\end{eqnarray}
where, in the right-hand side, one needs to substitute $\cos\theta=Z$. By analogy with the scalar case, we may identify the difference of the conditional probabilities with the phase shift between the fluctuating vector fields evaluated at a given geometric angle:
\begin{eqnarray}
p\left(+1\Big||\cos\theta|\right)-p\left(-1\Big||\cos\theta|\right)=\frac{2\tilde{\phi_0}}{\pi}\Big|_{|\cos\theta|}.\quad\quad
\end{eqnarray}
The remaining probability density, $p(|\cos\theta|)$, describes their geometric correlation.

\begin{figure}
\centering 
\includegraphics[width=0.99\columnwidth]{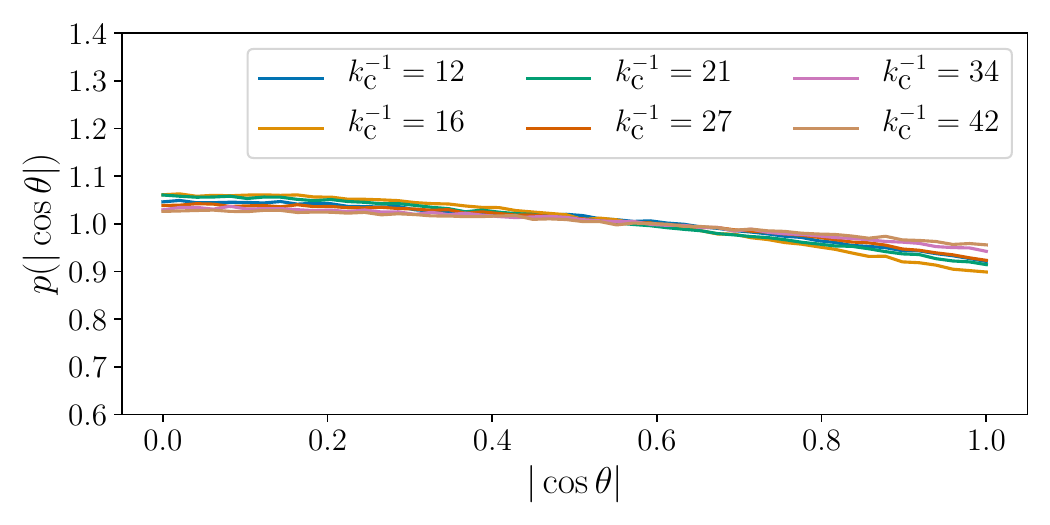}
\caption{Probability density function of the absolute value of the geometric angle between band-pass-filtered fluctuations of velocity and vorticity. No significant dependence on scale is observed. Data obtained from simulation A3.}
\label{fig:1536_Eh_05_PDF_ABS_updated}
\end{figure}

\begin{figure}
\centering 
\includegraphics[width=0.99\columnwidth]{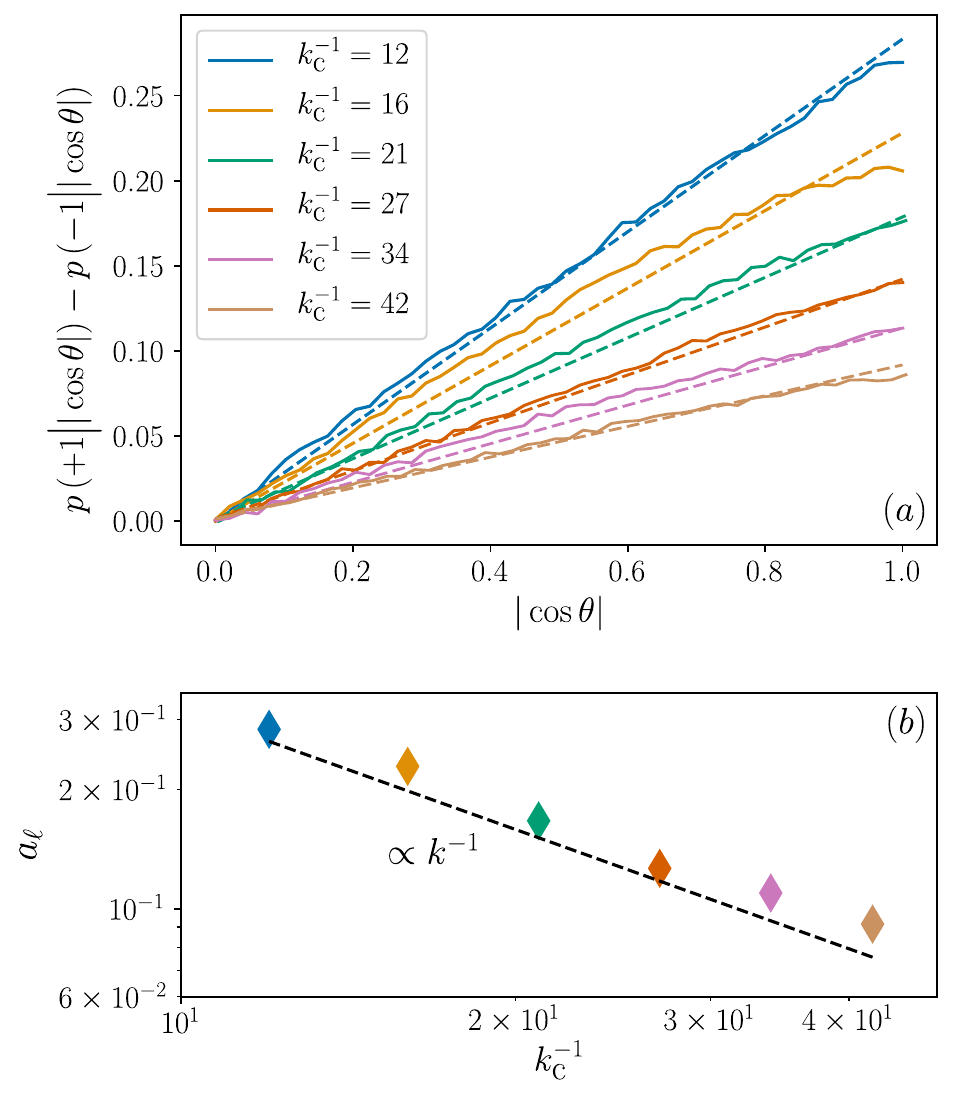}
\caption{(a) Difference between conditional probability of positive and negative alignment for a given $|\cos \theta|$ (solid lines) and linear fit of the data for each wavenumber range (dashed lines). The figure illustrates that, at each scale, the function $g(|\cos\theta|)$ is well approximated by a linear expression, i.e., $g(|\cos\theta|) \approx |\cos\theta|$. (b) Slope coefficients of the linear fits, $a_{\ell}$,  for the different wavenumber ranges. The color coding highlights the correspondence between data presented in subplots (a) and (b). Data obtained from simulation A3.}
\label{fig:linear_fit_figure}
\end{figure}

Using the probability density function in Eq.~(\ref{ProbaZ}), we may now average the $Z$-field:
\begin{eqnarray}
\langle Z\rangle=\int\limits_0^1\left( \frac{2\tilde{\phi_0}}{\pi}\Big|_{|\cos\theta|}\right) p(|\cos\theta|) \cos\theta\,d(\cos\theta).\quad
\end{eqnarray}
The first term in the integrand describes the contribution from the phase alignment between the velocity and vorticity fields, while the second term, $p(|\cos\theta|)$, describes the contribution from their geometric alignment. In principle, both terms may depend on the filtering scale $\ell$. Our numerical simulations, however, demonstrate that, quite crucially, the scale dependence factors out in the first term; that is,
\begin{eqnarray}
\frac{2\tilde{\phi_0}}{\pi}\Big|_{|\cos\theta|} \approx a_{\ell}g(|\cos\theta|) \approx a_{\ell}|\cos\theta|,
\end{eqnarray}
while the probability density of the geometric angle, $p(|\cos(\theta)|)$, turns out to be virtually independent of the scale.~Figs.~\ref{fig:1536_Eh_05_PDF_ABS_updated} and \ref{fig:linear_fit_figure}, obtained from time-averaged data from simulation A3, show that indeed the functions $p(|\cos\theta|)$ and $g(|\cos\theta|)$ do not vary with scale, while the phase-alignment function scales as $a_{\ell} \sim \ell \sim k^{-1}$, in agreement with the Fourier-space analysis of the previous section.~These conclusions hold for other values of  $\mathcal{R}_\mathcal{H}$.
As shown in Fig.~\ref{fig:linear_fit_figure}a, $g(|\cos\theta|)$ is well approximated by a linear function, i.e., $g(|\cos\theta|) \approx |\cos\theta|$, for all wavenumber ranges, with $a_{\ell}$ representing the slope of the linear fit at each scale (Fig.~\ref{fig:linear_fit_figure}b).~Our analysis demonstrates that it is phase alignment, not geometric alignment, that determines the scale-dependent reduction of helicity, so that
\begin{eqnarray}
\langle Z\rangle\approx C a_{\ell} \sim C a_0\ell/\ell_{0},
\end{eqnarray}
where we used $a_{\ell}\sim a_0(\ell/\ell_0)$, with $\ell_0$ the outer scale of turbulence, $a_0$ a constant dependent on net helicity in the system, and $C$ a scale-\textit{independent} constant:

\begin{eqnarray}
C=\int\limits_0^1 g(|\cos\theta|)p(|\cos\theta|)\cos\theta\,d(\cos\theta) \approx 1/3,\quad
\end{eqnarray}
where the last approximate equality can be obtained for $g(|\cos\theta|)\approx |\cos\theta|$ and $p(|\cos\theta|)\approx 1$.

\textit{Discussion and conclusions.}~\lucio{In this Letter, we demonstrate that the direct cascades of energy and helicity in Navier-Stokes turbulence can coexist because of a scale-dependence of the average Fourier phase angle between the fluctuations of velocity and vorticity.}~This behavior, termed \textit{dynamic phase alignment}, constitutes an essential mechanism for such joint direct cascade because the scalings $v_{\lambda} \sim k^{-1/3}$ and $\omega_{\lambda} \sim k v_{\lambda}$ are preserved in the presence of net helicity.~\lucio{We can write the helicity spectral scaling as $\mathcal{H}(k) \sim k^{-1} v_{\lambda}\omega_{\lambda}\cos \alpha_k \sim k^{-2/3} \cos \alpha_k$.~Deviations from $\mathcal{H}(k) \sim k^{-2/3}$ are underpinned by a dependence on scale of $\cos \alpha_k$, allowing energy and helicity to cascade forward while preserving conservation of both invariants.~We show that the observed spectrum $\mathcal{H}(k) \sim k^{-5/3}$ results from the scaling $\cos \alpha_k \propto k^{-1}$, which underlies the progressive balancing of turbulence (restoration of mirror symmetry) in the inertial range.}~We also demonstrate that there exists no significant scale-dependent \textit{geometric} alignment between velocity and vorticity, supporting our conclusion that \textit{phase} alignment is the primary mechanism contributing to the scale-dependent reduction of helicity.
In Ref.~\cite{Milanese2020DynamicTurbulence}, it was found that dynamic phase alignment between the fluctuations of electric and magnetic potentials underpins the joint forward cascade of energy and generalized kinetic helicity in a range of anisotropic turbulent plasma systems.~This remarkable similarity between systems governed by different sets of equations is suggestive of a powerful unifying paradigm, whereby the conservation of energy and a second invariant in a joint cascade determines the scale-dependent phase alignment between the two fields in the integrand of the second invariant. Dynamic phase alignment thus acquires importance as a mechanism regulating the dynamics in the presence of two invariants, arising from their conservation in the joint direct cascade, regardless of the details of the physical interactions. 

While we provided significant evidence for the existence and importance of dynamic phase alignment, we stopped short of characterizing how it emerges from the nonlinear interactions.~Dynamic phase alignment need not be regarded as necessarily alternative to other proposed paradigms for characterizing imbalanced turbulence \cite{Chen2003IntermittencyHelicityc, Chen2003TheTurbulence, Li2010GeometricalTurbulences, Alexakis2017HelicallyTurbulence, Alexakis2018CascadesFlowsb, Linkmann2018EffectsTurbulence, Buzzicotti2018EnergyRotation, Biferale2019HelicoidalInjection, Yan2020ScaleSpace, Yan2020DualFlows}.~As an example, Ref.~\cite{Chen2003TheTurbulence} explains the spectral scaling of helicity as arising from an imbalanced transfer between modes of positive and negative chirality within the helical decomposition framework \cite{Waleffe1992TheTurbulence}.~It is possible that dynamic phase alignment results from such transfer. 

Work supported by DoE grant No.~DE-FG02-91ER54109 (L.M.M. and N.F.L.), NSF CAREER award No. PHY-1654168 (N.F.L.), the Prof. Amar G. Bose Research Fellows Program at MIT (L.M.M. and N.F.L.), the Manson Benedict Fellowship of the MIT Nuclear Science and Engineering Department (L.M.M.) and NSF grants No. PHY-1707272 and PHY-2010098, NASA grant No. 80NSSC18K0646, and DOE grant No.
DESC0018266 (S.B.). We thank S. Chatterjee, A. Teimurazov, S. Sadhukhan, M.K. Sharma, R. Samtaney and M. Verma of the \texttt{Tarang} collaboration for support and helpful discussions. This research used resources  of the facilities of the Massachusetts Green High-Performance Computing Center (MGHPCC) and of the National Energy Research Scientific Computing Center (NERSC), a U.S. Department of Energy Office of Science User Facility operated under Contract No. DE-AC02-05CH11231. We thank Z. Zhao of the NERSC staff for outstanding technical support. 

\bibliography{references_PRL2}

\begin{thebibliography}{52}%
\makeatletter
\providecommand \@ifxundefined [1]{%
 \@ifx{#1\undefined}
}%
\providecommand \@ifnum [1]{%
 \ifnum #1\expandafter \@firstoftwo
 \else \expandafter \@secondoftwo
 \fi
}%
\providecommand \@ifx [1]{%
 \ifx #1\expandafter \@firstoftwo
 \else \expandafter \@secondoftwo
 \fi
}%
\providecommand \natexlab [1]{#1}%
\providecommand \enquote  [1]{``#1''}%
\providecommand \bibnamefont  [1]{#1}%
\providecommand \bibfnamefont [1]{#1}%
\providecommand \citenamefont [1]{#1}%
\providecommand \href@noop [0]{\@secondoftwo}%
\providecommand \href [0]{\begingroup \@sanitize@url \@href}%
\providecommand \@href[1]{\@@startlink{#1}\@@href}%
\providecommand \@@href[1]{\endgroup#1\@@endlink}%
\providecommand \@sanitize@url [0]{\catcode `\\12\catcode `\$12\catcode
  `\&12\catcode `\#12\catcode `\^12\catcode `\_12\catcode `\%12\relax}%
\providecommand \@@startlink[1]{}%
\providecommand \@@endlink[0]{}%
\providecommand \url  [0]{\begingroup\@sanitize@url \@url }%
\providecommand \@url [1]{\endgroup\@href {#1}{\urlprefix }}%
\providecommand \urlprefix  [0]{URL }%
\providecommand \Eprint [0]{\href }%
\providecommand \doibase [0]{https://doi.org/}%
\providecommand \selectlanguage [0]{\@gobble}%
\providecommand \bibinfo  [0]{\@secondoftwo}%
\providecommand \bibfield  [0]{\@secondoftwo}%
\providecommand \translation [1]{[#1]}%
\providecommand \BibitemOpen [0]{}%
\providecommand \bibitemStop [0]{}%
\providecommand \bibitemNoStop [0]{.\EOS\space}%
\providecommand \EOS [0]{\spacefactor3000\relax}%
\providecommand \BibitemShut  [1]{\csname bibitem#1\endcsname}%
\let\auto@bib@innerbib\@empty
\bibitem [{\citenamefont {Foias}\ \emph {et~al.}(2001)\citenamefont {Foias},
  \citenamefont {Manley}, \citenamefont {Rosa},\ and\ \citenamefont
  {Temam}}]{Foias2001Navier-StokesTurbulence}%
  \BibitemOpen
  \bibfield  {author} {\bibinfo {author} {\bibfnamefont {C.}~\bibnamefont
  {Foias}}, \bibinfo {author} {\bibfnamefont {O.}~\bibnamefont {Manley}},
  \bibinfo {author} {\bibfnamefont {R.}~\bibnamefont {Rosa}},\ and\ \bibinfo
  {author} {\bibfnamefont {R.}~\bibnamefont {Temam}},\ }\href
  {https://doi.org/10.1063/1.1522171} {\emph {\bibinfo {title} {Encyclopedia of
  Mathematics and its Applications}}}\ (\bibinfo  {publisher} {Cambridge
  University Press},\ \bibinfo {year} {2001})\BibitemShut {NoStop}%
\bibitem [{\citenamefont {Frisch}(1995)}]{Frisch1995Turbulence}%
  \BibitemOpen
  \bibfield  {author} {\bibinfo {author} {\bibfnamefont {U.}~\bibnamefont
  {Frisch}},\ }\href {https://doi.org/10.1017/CBO9781139170666} {\emph
  {\bibinfo {title} {{Turbulence}}}}\ (\bibinfo  {publisher} {Cambridge
  University Press},\ \bibinfo {year} {1995})\BibitemShut {NoStop}%
\bibitem [{\citenamefont {Kolmogorov}(1941)}]{Kolmogorov1941TheNumber}%
  \BibitemOpen
  \bibfield  {author} {\bibinfo {author} {\bibfnamefont {A.~N.}\ \bibnamefont
  {Kolmogorov}},\ }\bibfield  {title} {\bibinfo {title} {{The local structure
  of isotropic turbulence in an incompressible viscous fluid for very large
  Reynolds number}},\ }\href@noop {} {\bibfield  {journal} {\bibinfo  {journal}
  {Doklady Akademii nauk SSSR}\ }\textbf {\bibinfo {volume} {30}},\ \bibinfo
  {pages} {299} (\bibinfo {year} {1941})}\BibitemShut {NoStop}%
\bibitem [{\citenamefont {Moreau}(1961)}]{Moreau1961ConstantesBarotrope}%
  \BibitemOpen
  \bibfield  {author} {\bibinfo {author} {\bibfnamefont {J.~J.}\ \bibnamefont
  {Moreau}},\ }\bibfield  {title} {\bibinfo {title} {{Constantes d’un
  {\^{i}}lot tourbillonnaire en fluide parfait barotrope}},\ }\href
  {https://hal.archives-ouvertes.fr/hal-01865239/document} {\bibfield
  {journal} {\bibinfo  {journal} {Comptes rendus hebdomadaires des
  s{\'{e}}ances de l’Acad{\'{e}}mie des sciences}\ }\textbf {\bibinfo
  {volume} {252}},\ \bibinfo {pages} {2810} (\bibinfo {year}
  {1961})}\BibitemShut {NoStop}%
\bibitem [{\citenamefont {Moffatt}(1969)}]{Moffatt1969TheLines}%
  \BibitemOpen
  \bibfield  {author} {\bibinfo {author} {\bibfnamefont {H.~K.}\ \bibnamefont
  {Moffatt}},\ }\bibfield  {title} {\bibinfo {title} {{The degree of
  knottedness of tangled vortex lines}},\ }\href
  {https://doi.org/10.1017/S0022112069000991} {\bibfield  {journal} {\bibinfo
  {journal} {Journal of Fluid Mechanics}\ }\textbf {\bibinfo {volume} {35}},\
  \bibinfo {pages} {117} (\bibinfo {year} {1969})}\BibitemShut {NoStop}%
\bibitem [{\citenamefont {Chen}\ \emph
  {et~al.}(2003{\natexlab{a}})\citenamefont {Chen}, \citenamefont {Chen},
  \citenamefont {Eyink},\ and\ \citenamefont
  {Holm}}]{Chen2003IntermittencyHelicityc}%
  \BibitemOpen
  \bibfield  {author} {\bibinfo {author} {\bibfnamefont {Q.}~\bibnamefont
  {Chen}}, \bibinfo {author} {\bibfnamefont {S.}~\bibnamefont {Chen}}, \bibinfo
  {author} {\bibfnamefont {G.~L.}\ \bibnamefont {Eyink}},\ and\ \bibinfo
  {author} {\bibfnamefont {D.~D.}\ \bibnamefont {Holm}},\ }\bibfield  {title}
  {\bibinfo {title} {{Intermittency in the Joint Cascade of Energy and
  Helicity}},\ }\href {https://doi.org/10.1103/PhysRevLett.90.214503}
  {\bibfield  {journal} {\bibinfo  {journal} {Physical Review Letters}\
  }\textbf {\bibinfo {volume} {90}},\ \bibinfo {pages} {4} (\bibinfo {year}
  {2003}{\natexlab{a}})}\BibitemShut {NoStop}%
\bibitem [{\citenamefont {Alexakis}\ and\ \citenamefont
  {Biferale}(2018)}]{Alexakis2018CascadesFlowsb}%
  \BibitemOpen
  \bibfield  {author} {\bibinfo {author} {\bibfnamefont {A.}~\bibnamefont
  {Alexakis}}\ and\ \bibinfo {author} {\bibfnamefont {L.}~\bibnamefont
  {Biferale}},\ }\bibfield  {title} {\bibinfo {title} {{Cascades and
  transitions in turbulent flows}},\ }\href
  {https://doi.org/10.1016/j.physrep.2018.08.001} {\bibfield  {journal}
  {\bibinfo  {journal} {Physics Reports}\ }\textbf {\bibinfo {volume}
  {767-769}},\ \bibinfo {pages} {1} (\bibinfo {year} {2018})}\BibitemShut
  {NoStop}%
\bibitem [{\citenamefont {Pouquet}\ \emph
  {et~al.}(2019{\natexlab{a}})\citenamefont {Pouquet}, \citenamefont
  {Rosenberg}, \citenamefont {Stawarz},\ and\ \citenamefont
  {Marino}}]{Pouquet2019HelicityReviewb}%
  \BibitemOpen
  \bibfield  {author} {\bibinfo {author} {\bibfnamefont {A.}~\bibnamefont
  {Pouquet}}, \bibinfo {author} {\bibfnamefont {D.}~\bibnamefont {Rosenberg}},
  \bibinfo {author} {\bibfnamefont {J.~E.}\ \bibnamefont {Stawarz}},\ and\
  \bibinfo {author} {\bibfnamefont {R.}~\bibnamefont {Marino}},\ }\bibfield
  {title} {\bibinfo {title} {{Helicity Dynamics, Inverse, and Bidirectional
  Cascades in Fluid and Magnetohydrodynamic Turbulence: A Brief Review}},\
  }\href {https://doi.org/10.1029/2018EA000432} {\bibfield  {journal} {\bibinfo
   {journal} {Earth and Space Science}\ }\textbf {\bibinfo {volume} {6}},\
  \bibinfo {pages} {351} (\bibinfo {year} {2019}{\natexlab{a}})}\BibitemShut
  {NoStop}%
\bibitem [{\citenamefont {Brissaud}\ \emph {et~al.}(1973)\citenamefont
  {Brissaud}, \citenamefont {Frisch}, \citenamefont {Leorat}, \citenamefont
  {Lesieur},\ and\ \citenamefont {Mazure}}]{Brissaud1973HelicityTurbulence}%
  \BibitemOpen
  \bibfield  {author} {\bibinfo {author} {\bibfnamefont {A.}~\bibnamefont
  {Brissaud}}, \bibinfo {author} {\bibfnamefont {U.}~\bibnamefont {Frisch}},
  \bibinfo {author} {\bibfnamefont {J.}~\bibnamefont {Leorat}}, \bibinfo
  {author} {\bibfnamefont {M.}~\bibnamefont {Lesieur}},\ and\ \bibinfo {author}
  {\bibfnamefont {A.}~\bibnamefont {Mazure}},\ }\bibfield  {title} {\bibinfo
  {title} {{Helicity cascades in fully developed isotropic turbulence}},\
  }\href {https://doi.org/10.1063/1.1694520} {\bibfield  {journal} {\bibinfo
  {journal} {Physics of Fluids}\ }\textbf {\bibinfo {volume} {16}},\ \bibinfo
  {pages} {1366} (\bibinfo {year} {1973})}\BibitemShut {NoStop}%
\bibitem [{\citenamefont {Kraichnan}(1967)}]{Kraichnan1967InertialTurbulence}%
  \BibitemOpen
  \bibfield  {author} {\bibinfo {author} {\bibfnamefont {R.~H.}\ \bibnamefont
  {Kraichnan}},\ }\bibfield  {title} {\bibinfo {title} {{Inertial ranges in
  two-dimensional turbulence}},\ }\href {https://doi.org/10.1063/1.1762301}
  {\bibfield  {journal} {\bibinfo  {journal} {Physics of Fluids}\ }\textbf
  {\bibinfo {volume} {10}},\ \bibinfo {pages} {1417} (\bibinfo {year}
  {1967})}\BibitemShut {NoStop}%
\bibitem [{\citenamefont {Frisch}\ \emph {et~al.}(1975)\citenamefont {Frisch},
  \citenamefont {Pouquet}, \citenamefont {L{\'{e}}orat},\ and\ \citenamefont
  {Mazure}}]{Frisch1975PossibilityTurbulence}%
  \BibitemOpen
  \bibfield  {author} {\bibinfo {author} {\bibfnamefont {U.}~\bibnamefont
  {Frisch}}, \bibinfo {author} {\bibfnamefont {A.}~\bibnamefont {Pouquet}},
  \bibinfo {author} {\bibfnamefont {J.}~\bibnamefont {L{\'{e}}orat}},\ and\
  \bibinfo {author} {\bibfnamefont {A.}~\bibnamefont {Mazure}},\ }\bibfield
  {title} {\bibinfo {title} {{Possibility of an inverse cascade of magnetic
  helicity in magnetohydrodynamic turbulence}},\ }\href
  {https://doi.org/10.1017/S002211207500122X} {\bibfield  {journal} {\bibinfo
  {journal} {Journal of Fluid Mechanics}\ }\textbf {\bibinfo {volume} {68}},\
  \bibinfo {pages} {769} (\bibinfo {year} {1975})}\BibitemShut {NoStop}%
\bibitem [{\citenamefont
  {Hasegawa}(1985)}]{Hasegawa1985Self-organizationMedia}%
  \BibitemOpen
  \bibfield  {author} {\bibinfo {author} {\bibfnamefont {A.}~\bibnamefont
  {Hasegawa}},\ }\bibfield  {title} {\bibinfo {title} {{Self-organization
  processes in continuous media}},\ }\href
  {https://doi.org/10.1080/00018738500101721} {\bibfield  {journal} {\bibinfo
  {journal} {Advances in Physics}\ }\textbf {\bibinfo {volume} {34}},\ \bibinfo
  {pages} {1} (\bibinfo {year} {1985})}\BibitemShut {NoStop}%
\bibitem [{\citenamefont {Zakharov}\ \emph {et~al.}(1992)\citenamefont
  {Zakharov}, \citenamefont {Lvov},\ and\ \citenamefont
  {Falkovich}}]{Zakharov1992KolmogorovTurbulence}%
  \BibitemOpen
  \bibfield  {author} {\bibinfo {author} {\bibfnamefont {V.~E.}\ \bibnamefont
  {Zakharov}}, \bibinfo {author} {\bibfnamefont {V.~S.}\ \bibnamefont {Lvov}},\
  and\ \bibinfo {author} {\bibfnamefont {G.}~\bibnamefont {Falkovich}},\ }\href
  {https://doi.org/10.1007/978-3-642-50052-7} {\emph {\bibinfo {title}
  {Springer series in nonlinear dynamics}}},\ Springer Series in Nonlinear
  Dynamics\ (\bibinfo  {publisher} {Springer Berlin Heidelberg},\ \bibinfo
  {address} {Berlin, Heidelberg},\ \bibinfo {year} {1992})\BibitemShut
  {NoStop}%
\bibitem [{\citenamefont {Kraichnan}(1973)}]{Kraichnan1973HelicalEquilibrium}%
  \BibitemOpen
  \bibfield  {author} {\bibinfo {author} {\bibfnamefont {R.~H.}\ \bibnamefont
  {Kraichnan}},\ }\bibfield  {title} {\bibinfo {title} {{Helical turbulence and
  absolute equilibrium}},\ }\href {https://doi.org/10.1017/S0022112073001837}
  {\bibfield  {journal} {\bibinfo  {journal} {Journal of Fluid Mechanics}\
  }\textbf {\bibinfo {volume} {59}},\ \bibinfo {pages} {745} (\bibinfo {year}
  {1973})}\BibitemShut {NoStop}%
\bibitem [{\citenamefont {Borue}\ and\ \citenamefont
  {Orszag}(1997)}]{Borue1997SpectraTurbulence}%
  \BibitemOpen
  \bibfield  {author} {\bibinfo {author} {\bibfnamefont {V.}~\bibnamefont
  {Borue}}\ and\ \bibinfo {author} {\bibfnamefont {S.~A.}\ \bibnamefont
  {Orszag}},\ }\bibfield  {title} {\bibinfo {title} {{Spectra in helical
  three-dimensional homogeneous isotropic turbulence}},\ }\href
  {https://doi.org/10.1103/PhysRevE.55.7005} {\bibfield  {journal} {\bibinfo
  {journal} {Physical Review E - Statistical Physics, Plasmas, Fluids, and
  Related Interdisciplinary Topics}\ }\textbf {\bibinfo {volume} {55}},\
  \bibinfo {pages} {7005} (\bibinfo {year} {1997})}\BibitemShut {NoStop}%
\bibitem [{\citenamefont {Koprov}\ \emph {et~al.}(2005)\citenamefont {Koprov},
  \citenamefont {Koprov}, \citenamefont {Ponomarev},\ and\ \citenamefont
  {Chkhetiani}}]{Koprov2005ExperimentalLayer}%
  \BibitemOpen
  \bibfield  {author} {\bibinfo {author} {\bibfnamefont {B.~M.}\ \bibnamefont
  {Koprov}}, \bibinfo {author} {\bibfnamefont {V.~M.}\ \bibnamefont {Koprov}},
  \bibinfo {author} {\bibfnamefont {V.~M.}\ \bibnamefont {Ponomarev}},\ and\
  \bibinfo {author} {\bibfnamefont {O.~G.}\ \bibnamefont {Chkhetiani}},\
  }\bibfield  {title} {\bibinfo {title} {{Experimental studies of turbulent
  helicity and its spectrum in the atmospheric boundary layer}},\ }\href
  {https://doi.org/10.1134/1.2039983} {\bibfield  {journal} {\bibinfo
  {journal} {Doklady Physics}\ }\textbf {\bibinfo {volume} {50}},\ \bibinfo
  {pages} {419} (\bibinfo {year} {2005})}\BibitemShut {NoStop}%
\bibitem [{\citenamefont {Qu}\ \emph {et~al.}(2018)\citenamefont {Qu},
  \citenamefont {Naso},\ and\ \citenamefont {Bos}}]{Qu2018CascadesTurbulence}%
  \BibitemOpen
  \bibfield  {author} {\bibinfo {author} {\bibfnamefont {B.}~\bibnamefont
  {Qu}}, \bibinfo {author} {\bibfnamefont {A.}~\bibnamefont {Naso}},\ and\
  \bibinfo {author} {\bibfnamefont {W.~J.}\ \bibnamefont {Bos}},\ }\bibfield
  {title} {\bibinfo {title} {{Cascades of energy and helicity in axisymmetric
  turbulence}},\ }\href {https://doi.org/10.1103/PhysRevFluids.3.014607}
  {\bibfield  {journal} {\bibinfo  {journal} {Physical Review Fluids}\ }\textbf
  {\bibinfo {volume} {3}},\ \bibinfo {pages} {1} (\bibinfo {year}
  {2018})}\BibitemShut {NoStop}%
\bibitem [{\citenamefont {Pouquet}\ \emph
  {et~al.}(2019{\natexlab{b}})\citenamefont {Pouquet}, \citenamefont
  {Rosenberg}, \citenamefont {Stawarz},\ and\ \citenamefont
  {Marino}}]{Pouquet2019HelicityReview}%
  \BibitemOpen
  \bibfield  {author} {\bibinfo {author} {\bibfnamefont {A.}~\bibnamefont
  {Pouquet}}, \bibinfo {author} {\bibfnamefont {D.}~\bibnamefont {Rosenberg}},
  \bibinfo {author} {\bibfnamefont {J.~E.}\ \bibnamefont {Stawarz}},\ and\
  \bibinfo {author} {\bibfnamefont {R.}~\bibnamefont {Marino}},\ }\bibfield
  {title} {\bibinfo {title} {{Helicity Dynamics, Inverse, and Bidirectional
  Cascades in Fluid and Magnetohydrodynamic Turbulence: A Brief Review}},\
  }\href {https://doi.org/10.1029/2018EA000432} {\bibfield  {journal} {\bibinfo
   {journal} {Earth and Space Science}\ }\textbf {\bibinfo {volume} {6}},\
  \bibinfo {pages} {351} (\bibinfo {year} {2019}{\natexlab{b}})}\BibitemShut
  {NoStop}%
\bibitem [{\citenamefont {Plunian}\ \emph {et~al.}(2020)\citenamefont
  {Plunian}, \citenamefont {Teimurazov}, \citenamefont {Stepanov},\ and\
  \citenamefont {Verma}}]{Plunian2020InverseTurbulenceb}%
  \BibitemOpen
  \bibfield  {author} {\bibinfo {author} {\bibfnamefont {F.}~\bibnamefont
  {Plunian}}, \bibinfo {author} {\bibfnamefont {A.}~\bibnamefont {Teimurazov}},
  \bibinfo {author} {\bibfnamefont {R.}~\bibnamefont {Stepanov}},\ and\
  \bibinfo {author} {\bibfnamefont {M.~K.}\ \bibnamefont {Verma}},\ }\bibfield
  {title} {\bibinfo {title} {{Inverse cascade of energy in helical
  turbulence}},\ }\bibfield  {journal} {\bibinfo  {journal} {Journal of Fluid
  Mechanics}\ }\textbf {\bibinfo {volume} {895}},\ \href
  {https://doi.org/10.1017/jfm.2020.307} {10.1017/jfm.2020.307} (\bibinfo
  {year} {2020})\BibitemShut {NoStop}%
\bibitem [{\citenamefont {Chen}\ \emph
  {et~al.}(2003{\natexlab{b}})\citenamefont {Chen}, \citenamefont {Chen},\ and\
  \citenamefont {Eyink}}]{Chen2003TheTurbulence}%
  \BibitemOpen
  \bibfield  {author} {\bibinfo {author} {\bibfnamefont {Q.}~\bibnamefont
  {Chen}}, \bibinfo {author} {\bibfnamefont {S.}~\bibnamefont {Chen}},\ and\
  \bibinfo {author} {\bibfnamefont {G.~L.}\ \bibnamefont {Eyink}},\ }\bibfield
  {title} {\bibinfo {title} {{The joint cascade of energy and helicity in
  three-dimensional turbulence}},\ }\href {https://doi.org/10.1063/1.1533070}
  {\bibfield  {journal} {\bibinfo  {journal} {Physics of Fluids}\ }\textbf
  {\bibinfo {volume} {15}},\ \bibinfo {pages} {361} (\bibinfo {year}
  {2003}{\natexlab{b}})}\BibitemShut {NoStop}%
\bibitem [{\citenamefont {Eyink}(2005)}]{Eyink2005LocalityCascades}%
  \BibitemOpen
  \bibfield  {author} {\bibinfo {author} {\bibfnamefont {G.~L.}\ \bibnamefont
  {Eyink}},\ }\bibfield  {title} {\bibinfo {title} {{Locality of turbulent
  cascades}},\ }\href {https://doi.org/10.1016/j.physd.2005.05.018} {\bibfield
  {journal} {\bibinfo  {journal} {Physica D: Nonlinear Phenomena}\ }\textbf
  {\bibinfo {volume} {207}},\ \bibinfo {pages} {91} (\bibinfo {year}
  {2005})}\BibitemShut {NoStop}%
\bibitem [{\citenamefont {Baerenzung}\ \emph {et~al.}(2008)\citenamefont
  {Baerenzung}, \citenamefont {Politano}, \citenamefont {Ponty},\ and\
  \citenamefont {Pouquet}}]{Baerenzung2008SpectralHelicity}%
  \BibitemOpen
  \bibfield  {author} {\bibinfo {author} {\bibfnamefont {J.}~\bibnamefont
  {Baerenzung}}, \bibinfo {author} {\bibfnamefont {H.}~\bibnamefont
  {Politano}}, \bibinfo {author} {\bibfnamefont {Y.}~\bibnamefont {Ponty}},\
  and\ \bibinfo {author} {\bibfnamefont {A.}~\bibnamefont {Pouquet}},\
  }\bibfield  {title} {\bibinfo {title} {{Spectral modeling of turbulent flows
  and the role of helicity}},\ }\href
  {https://doi.org/10.1103/PhysRevE.77.046303} {\bibfield  {journal} {\bibinfo
  {journal} {Physical Review E - Statistical, Nonlinear, and Soft Matter
  Physics}\ }\textbf {\bibinfo {volume} {77}},\ \bibinfo {pages} {046303}
  (\bibinfo {year} {2008})}\BibitemShut {NoStop}%
\bibitem [{\citenamefont {Stepanov}\ \emph {et~al.}(2009)\citenamefont
  {Stepanov}, \citenamefont {Frik},\ and\ \citenamefont
  {Shestakov}}]{Stepanov2009SpectralTurbulence}%
  \BibitemOpen
  \bibfield  {author} {\bibinfo {author} {\bibfnamefont {R.~A.}\ \bibnamefont
  {Stepanov}}, \bibinfo {author} {\bibfnamefont {P.~G.}\ \bibnamefont {Frik}},\
  and\ \bibinfo {author} {\bibfnamefont {A.~V.}\ \bibnamefont {Shestakov}},\
  }\bibfield  {title} {\bibinfo {title} {{Spectral properties of helical
  turbulence}},\ }\href {https://doi.org/10.1134/S0015462809050044} {\bibfield
  {journal} {\bibinfo  {journal} {Fluid Dynamics}\ }\textbf {\bibinfo {volume}
  {44}},\ \bibinfo {pages} {658} (\bibinfo {year} {2009})}\BibitemShut
  {NoStop}%
\bibitem [{\citenamefont {Choi}\ \emph {et~al.}(2009)\citenamefont {Choi},
  \citenamefont {Kim},\ and\ \citenamefont
  {Lee}}]{Choi2009AlignmentTurbulence}%
  \BibitemOpen
  \bibfield  {author} {\bibinfo {author} {\bibfnamefont {Y.}~\bibnamefont
  {Choi}}, \bibinfo {author} {\bibfnamefont {B.-G.}\ \bibnamefont {Kim}},\ and\
  \bibinfo {author} {\bibfnamefont {C.}~\bibnamefont {Lee}},\ }\bibfield
  {title} {\bibinfo {title} {{Alignment of velocity and vorticity and the
  intermittent distribution of helicity in isotropic turbulence}},\ }\bibfield
  {journal} {\bibinfo  {journal} {Physical Review E}\ }\textbf {\bibinfo
  {volume} {80}},\ \href {https://doi.org/10.1103/physreve.80.017301}
  {10.1103/physreve.80.017301} (\bibinfo {year} {2009})\BibitemShut {NoStop}%
\bibitem [{\citenamefont {Teitelbaum}\ and\ \citenamefont
  {Mininni}(2009)}]{Teitelbaum2009EffectFlows}%
  \BibitemOpen
  \bibfield  {author} {\bibinfo {author} {\bibfnamefont {T.}~\bibnamefont
  {Teitelbaum}}\ and\ \bibinfo {author} {\bibfnamefont {P.~D.}\ \bibnamefont
  {Mininni}},\ }\bibfield  {title} {\bibinfo {title} {{Effect of Helicity and
  Rotation on the Free Decay of Turbulent Flows}},\ }\href
  {https://doi.org/10.1103/PhysRevLett.103.014501} {\bibfield  {journal}
  {\bibinfo  {journal} {Physical Review Letters}\ }\textbf {\bibinfo {volume}
  {103}},\ \bibinfo {pages} {014501} (\bibinfo {year} {2009})}\BibitemShut
  {NoStop}%
\bibitem [{\citenamefont {Plunian}\ \emph {et~al.}(2011)\citenamefont
  {Plunian}, \citenamefont {Lessinnes}, \citenamefont {Carati},\ and\
  \citenamefont {Stepanov}}]{Plunian2011HelicityScalings}%
  \BibitemOpen
  \bibfield  {author} {\bibinfo {author} {\bibfnamefont {F.}~\bibnamefont
  {Plunian}}, \bibinfo {author} {\bibfnamefont {T.}~\bibnamefont {Lessinnes}},
  \bibinfo {author} {\bibfnamefont {D.}~\bibnamefont {Carati}},\ and\ \bibinfo
  {author} {\bibfnamefont {R.}~\bibnamefont {Stepanov}},\ }\bibfield  {title}
  {\bibinfo {title} {{Helicity scalings}},\ }in\ \href
  {https://doi.org/10.1088/1742-6596/318/4/042013} {\emph {\bibinfo {booktitle}
  {Journal of Physics: Conference Series}}},\ Vol.\ \bibinfo {volume} {318}\
  (\bibinfo  {publisher} {Institute of Physics Publishing},\ \bibinfo {year}
  {2011})\BibitemShut {NoStop}%
\bibitem [{\citenamefont {Pouquet}\ \emph {et~al.}(2013)\citenamefont
  {Pouquet}, \citenamefont {Sen}, \citenamefont {Rosenberg}, \citenamefont
  {Mininni},\ and\ \citenamefont {Baerenzung}}]{Pouquet2013InverseFlows}%
  \BibitemOpen
  \bibfield  {author} {\bibinfo {author} {\bibfnamefont {A.}~\bibnamefont
  {Pouquet}}, \bibinfo {author} {\bibfnamefont {A.}~\bibnamefont {Sen}},
  \bibinfo {author} {\bibfnamefont {D.}~\bibnamefont {Rosenberg}}, \bibinfo
  {author} {\bibfnamefont {P.~D.}\ \bibnamefont {Mininni}},\ and\ \bibinfo
  {author} {\bibfnamefont {J.}~\bibnamefont {Baerenzung}},\ }\bibfield  {title}
  {\bibinfo {title} {{Inverse cascades in turbulence and the case of rotating
  flows}},\ }\href {https://doi.org/10.1088/0031-8949/2013/T155/014032}
  {\bibfield  {journal} {\bibinfo  {journal} {Physica Scripta}\ }\textbf
  {\bibinfo {volume} {88}},\ \bibinfo {pages} {014032} (\bibinfo {year}
  {2013})}\BibitemShut {NoStop}%
\bibitem [{\citenamefont {Biferale}\ \emph {et~al.}(2013)\citenamefont
  {Biferale}, \citenamefont {Musacchio},\ and\ \citenamefont
  {Toschi}}]{Biferale2013SplitTurbulence}%
  \BibitemOpen
  \bibfield  {author} {\bibinfo {author} {\bibfnamefont {L.}~\bibnamefont
  {Biferale}}, \bibinfo {author} {\bibfnamefont {S.}~\bibnamefont
  {Musacchio}},\ and\ \bibinfo {author} {\bibfnamefont {F.}~\bibnamefont
  {Toschi}},\ }\bibfield  {title} {\bibinfo {title} {{Split energy-helicity
  cascades in three-dimensional homogeneous and isotropic turbulence}},\ }\href
  {https://doi.org/10.1017/jfm.2013.349} {\bibfield  {journal} {\bibinfo
  {journal} {Journal of Fluid Mechanics}\ }\textbf {\bibinfo {volume} {730}},\
  \bibinfo {pages} {309} (\bibinfo {year} {2013})}\BibitemShut {NoStop}%
\bibitem [{\citenamefont {Gledzer}\ and\ \citenamefont
  {Chkhetiani}(2015)}]{Gledzer2015InverseModes}%
  \BibitemOpen
  \bibfield  {author} {\bibinfo {author} {\bibfnamefont {E.~B.}\ \bibnamefont
  {Gledzer}}\ and\ \bibinfo {author} {\bibfnamefont {O.~G.}\ \bibnamefont
  {Chkhetiani}},\ }\bibfield  {title} {\bibinfo {title} {{Inverse energy
  cascade in developed turbulence at the breaking of the symmetry of helical
  modes}},\ }\href {https://doi.org/10.1134/S0021364015190066} {\bibfield
  {journal} {\bibinfo  {journal} {JETP Letters}\ }\textbf {\bibinfo {volume}
  {102}},\ \bibinfo {pages} {465} (\bibinfo {year} {2015})}\BibitemShut
  {NoStop}%
\bibitem [{\citenamefont {De~Pietro}\ \emph {et~al.}(2015)\citenamefont
  {De~Pietro}, \citenamefont {Biferale},\ and\ \citenamefont
  {Mailybaev}}]{DePietro2015InverseTurbulence}%
  \BibitemOpen
  \bibfield  {author} {\bibinfo {author} {\bibfnamefont {M.}~\bibnamefont
  {De~Pietro}}, \bibinfo {author} {\bibfnamefont {L.}~\bibnamefont
  {Biferale}},\ and\ \bibinfo {author} {\bibfnamefont {A.~A.}\ \bibnamefont
  {Mailybaev}},\ }\bibfield  {title} {\bibinfo {title} {{Inverse energy cascade
  in nonlocal helical shell models of turbulence}},\ }\href
  {https://doi.org/10.1103/PhysRevE.92.043021} {\bibfield  {journal} {\bibinfo
  {journal} {Physical Review E - Statistical, Nonlinear, and Soft Matter
  Physics}\ }\textbf {\bibinfo {volume} {92}},\ \bibinfo {pages} {043021}
  (\bibinfo {year} {2015})}\BibitemShut {NoStop}%
\bibitem [{\citenamefont {Sahoo}\ \emph {et~al.}(2015)\citenamefont {Sahoo},
  \citenamefont {Bonaccorso},\ and\ \citenamefont
  {Biferale}}]{Sahoo2015RoleFluctuations}%
  \BibitemOpen
  \bibfield  {author} {\bibinfo {author} {\bibfnamefont {G.}~\bibnamefont
  {Sahoo}}, \bibinfo {author} {\bibfnamefont {F.}~\bibnamefont {Bonaccorso}},\
  and\ \bibinfo {author} {\bibfnamefont {L.}~\bibnamefont {Biferale}},\
  }\bibfield  {title} {\bibinfo {title} {{Role of helicity for large- and
  small-scale turbulent fluctuations}},\ }\href
  {https://doi.org/10.1103/PhysRevE.92.051002} {\bibfield  {journal} {\bibinfo
  {journal} {Physical Review E - Statistical, Nonlinear, and Soft Matter
  Physics}\ }\textbf {\bibinfo {volume} {92}},\ \bibinfo {pages} {051002}
  (\bibinfo {year} {2015})}\BibitemShut {NoStop}%
\bibitem [{\citenamefont {Stepanov}\ \emph {et~al.}(2015)\citenamefont
  {Stepanov}, \citenamefont {Golbraikh}, \citenamefont {Frick},\ and\
  \citenamefont {Shestakov}}]{Stepanov2015HinderedTurbulence}%
  \BibitemOpen
  \bibfield  {author} {\bibinfo {author} {\bibfnamefont {R.}~\bibnamefont
  {Stepanov}}, \bibinfo {author} {\bibfnamefont {E.}~\bibnamefont {Golbraikh}},
  \bibinfo {author} {\bibfnamefont {P.}~\bibnamefont {Frick}},\ and\ \bibinfo
  {author} {\bibfnamefont {A.}~\bibnamefont {Shestakov}},\ }\bibfield  {title}
  {\bibinfo {title} {{Hindered Energy Cascade in Highly Helical Isotropic
  Turbulence}},\ }\href {https://doi.org/10.1103/PhysRevLett.115.234501}
  {\bibfield  {journal} {\bibinfo  {journal} {Physical Review Letters}\
  }\textbf {\bibinfo {volume} {115}},\ \bibinfo {pages} {234501} (\bibinfo
  {year} {2015})}\BibitemShut {NoStop}%
\bibitem [{\citenamefont {Imazio}\ and\ \citenamefont
  {Mininni}(2017)}]{Imazio2017PassiveTurbulence}%
  \BibitemOpen
  \bibfield  {author} {\bibinfo {author} {\bibfnamefont {P.~R.}\ \bibnamefont
  {Imazio}}\ and\ \bibinfo {author} {\bibfnamefont {P.~D.}\ \bibnamefont
  {Mininni}},\ }\bibfield  {title} {\bibinfo {title} {{Passive scalars: Mixing,
  diffusion, and intermittency in helical and nonhelical rotating
  turbulence}},\ }\href {https://doi.org/10.1103/PhysRevE.95.033103} {\bibfield
   {journal} {\bibinfo  {journal} {Physical Review E}\ }\textbf {\bibinfo
  {volume} {95}},\ \bibinfo {pages} {033103} (\bibinfo {year}
  {2017})}\BibitemShut {NoStop}%
\bibitem [{\citenamefont {Sahoo}\ \emph {et~al.}(2017)\citenamefont {Sahoo},
  \citenamefont {De~Pietro},\ and\ \citenamefont
  {Biferale}}]{Sahoo2017HelicityModels}%
  \BibitemOpen
  \bibfield  {author} {\bibinfo {author} {\bibfnamefont {G.}~\bibnamefont
  {Sahoo}}, \bibinfo {author} {\bibfnamefont {M.}~\bibnamefont {De~Pietro}},\
  and\ \bibinfo {author} {\bibfnamefont {L.}~\bibnamefont {Biferale}},\
  }\bibfield  {title} {\bibinfo {title} {{Helicity statistics in homogeneous
  and isotropic turbulence and turbulence models}},\ }\href
  {https://doi.org/10.1103/PhysRevFluids.2.024601} {\bibfield  {journal}
  {\bibinfo  {journal} {Physical Review Fluids}\ }\textbf {\bibinfo {volume}
  {2}},\ \bibinfo {pages} {24601} (\bibinfo {year} {2017})}\BibitemShut
  {NoStop}%
\bibitem [{\citenamefont {Chkhetiani}\ and\ \citenamefont
  {Gledzer}(2017)}]{Chkhetiani2017HelicalGeneration}%
  \BibitemOpen
  \bibfield  {author} {\bibinfo {author} {\bibfnamefont {O.~G.}\ \bibnamefont
  {Chkhetiani}}\ and\ \bibinfo {author} {\bibfnamefont {E.~B.}\ \bibnamefont
  {Gledzer}},\ }\bibfield  {title} {\bibinfo {title} {{Helical turbulence with
  small-scale energy and helicity sources and external intermediate scale
  noises as the origin of large scale generation}},\ }\href
  {https://doi.org/10.1016/j.physa.2017.05.027} {\bibfield  {journal} {\bibinfo
   {journal} {Physica A: Statistical Mechanics and its Applications}\ }\textbf
  {\bibinfo {volume} {486}},\ \bibinfo {pages} {416} (\bibinfo {year}
  {2017})}\BibitemShut {NoStop}%
\bibitem [{\citenamefont {Briard}\ \emph {et~al.}(2017)\citenamefont {Briard},
  \citenamefont {Biferale},\ and\ \citenamefont
  {Gomez}}]{Briard2017ClosureTurbulence}%
  \BibitemOpen
  \bibfield  {author} {\bibinfo {author} {\bibfnamefont {A.}~\bibnamefont
  {Briard}}, \bibinfo {author} {\bibfnamefont {L.}~\bibnamefont {Biferale}},\
  and\ \bibinfo {author} {\bibfnamefont {T.}~\bibnamefont {Gomez}},\ }\bibfield
   {title} {\bibinfo {title} {{Closure theory for the split energy-helicity
  cascades in homogeneous isotropic homochiral turbulence}},\ }\href
  {https://doi.org/10.1103/PhysRevFluids.2.102602} {\bibfield  {journal}
  {\bibinfo  {journal} {Physical Review Fluids}\ }\textbf {\bibinfo {volume}
  {2}},\ \bibinfo {pages} {102602} (\bibinfo {year} {2017})}\BibitemShut
  {NoStop}%
\bibitem [{\citenamefont {Teimurazov}\ \emph {et~al.}(2018)\citenamefont
  {Teimurazov}, \citenamefont {Stepanov}, \citenamefont {Verma}, \citenamefont
  {Barman}, \citenamefont {Kumar},\ and\ \citenamefont
  {Sadhukhan}}]{Teimurazov2018DirectCode}%
  \BibitemOpen
  \bibfield  {author} {\bibinfo {author} {\bibfnamefont {A.~S.}\ \bibnamefont
  {Teimurazov}}, \bibinfo {author} {\bibfnamefont {R.~A.}\ \bibnamefont
  {Stepanov}}, \bibinfo {author} {\bibfnamefont {M.~K.}\ \bibnamefont {Verma}},
  \bibinfo {author} {\bibfnamefont {S.}~\bibnamefont {Barman}}, \bibinfo
  {author} {\bibfnamefont {A.}~\bibnamefont {Kumar}},\ and\ \bibinfo {author}
  {\bibfnamefont {S.}~\bibnamefont {Sadhukhan}},\ }\bibfield  {title} {\bibinfo
  {title} {{Direct Numerical Simulation of Homogeneous Isotropic Helical
  Turbulence with the TARANG Code}},\ }\href
  {https://doi.org/10.1134/S0021894418070131} {\bibfield  {journal} {\bibinfo
  {journal} {Journal of Applied Mechanics and Technical Physics}\ }\textbf
  {\bibinfo {volume} {59}},\ \bibinfo {pages} {1279} (\bibinfo {year}
  {2018})}\BibitemShut {NoStop}%
\bibitem [{\citenamefont {Sahoo}\ and\ \citenamefont
  {Biferale}(2018)}]{Sahoo2018EnergyEquations}%
  \BibitemOpen
  \bibfield  {author} {\bibinfo {author} {\bibfnamefont {G.}~\bibnamefont
  {Sahoo}}\ and\ \bibinfo {author} {\bibfnamefont {L.}~\bibnamefont
  {Biferale}},\ }\bibfield  {title} {\bibinfo {title} {{Energy cascade and
  intermittency in helically decomposed Navier-Stokes equations}},\ }\href
  {https://doi.org/10.1088/1873-7005/aa839a} {\bibfield  {journal} {\bibinfo
  {journal} {Fluid Dynamics Research}\ }\textbf {\bibinfo {volume} {50}},\
  \bibinfo {pages} {011420} (\bibinfo {year} {2018})}\BibitemShut {NoStop}%
\bibitem [{\citenamefont {Yan}\ \emph {et~al.}(2020{\natexlab{a}})\citenamefont
  {Yan}, \citenamefont {Li},\ and\ \citenamefont {Yu}}]{Yan2020ScaleSpace}%
  \BibitemOpen
  \bibfield  {author} {\bibinfo {author} {\bibfnamefont {Z.}~\bibnamefont
  {Yan}}, \bibinfo {author} {\bibfnamefont {X.}~\bibnamefont {Li}},\ and\
  \bibinfo {author} {\bibfnamefont {C.}~\bibnamefont {Yu}},\ }\bibfield
  {title} {\bibinfo {title} {{Scale locality of helicity cascade in physical
  space}},\ }\href {https://doi.org/10.1063/5.0013009} {\bibfield  {journal}
  {\bibinfo  {journal} {Physics of Fluids}\ }\textbf {\bibinfo {volume} {32}},\
  \bibinfo {pages} {61705} (\bibinfo {year} {2020}{\natexlab{a}})}\BibitemShut
  {NoStop}%
\bibitem [{\citenamefont {Verma}\ \emph {et~al.}(2013)\citenamefont {Verma},
  \citenamefont {Chatterjee}, \citenamefont {Sandeep~Reddy}, \citenamefont
  {Yadav}, \citenamefont {Paul}, \citenamefont {Chandra},\ and\ \citenamefont
  {Samtaney}}]{Verma2013BenchmarkingSimulations}%
  \BibitemOpen
  \bibfield  {author} {\bibinfo {author} {\bibfnamefont {M.~K.}\ \bibnamefont
  {Verma}}, \bibinfo {author} {\bibfnamefont {A.}~\bibnamefont {Chatterjee}},
  \bibinfo {author} {\bibfnamefont {K.}~\bibnamefont {Sandeep~Reddy}}, \bibinfo
  {author} {\bibfnamefont {R.~K.}\ \bibnamefont {Yadav}}, \bibinfo {author}
  {\bibfnamefont {S.}~\bibnamefont {Paul}}, \bibinfo {author} {\bibfnamefont
  {M.}~\bibnamefont {Chandra}},\ and\ \bibinfo {author} {\bibfnamefont
  {R.}~\bibnamefont {Samtaney}},\ }\bibfield  {title} {\bibinfo {title}
  {{Benchmarking and scaling studies of pseudospectral code Tarang for
  turbulence simulations}},\ }\href {https://doi.org/10.1007/s12043-013-0594-4}
  {\bibfield  {journal} {\bibinfo  {journal} {Pramana - Journal of Physics}\
  }\textbf {\bibinfo {volume} {81}},\ \bibinfo {pages} {617} (\bibinfo {year}
  {2013})}\BibitemShut {NoStop}%
\bibitem [{\citenamefont {Chatterjee}\ \emph {et~al.}(2018)\citenamefont
  {Chatterjee}, \citenamefont {Verma}, \citenamefont {Kumar}, \citenamefont
  {Samtaney}, \citenamefont {Hadri},\ and\ \citenamefont
  {Khurram}}]{Chatterjee2018ScalingCoresb}%
  \BibitemOpen
  \bibfield  {author} {\bibinfo {author} {\bibfnamefont {A.~G.}\ \bibnamefont
  {Chatterjee}}, \bibinfo {author} {\bibfnamefont {M.~K.}\ \bibnamefont
  {Verma}}, \bibinfo {author} {\bibfnamefont {A.}~\bibnamefont {Kumar}},
  \bibinfo {author} {\bibfnamefont {R.}~\bibnamefont {Samtaney}}, \bibinfo
  {author} {\bibfnamefont {B.}~\bibnamefont {Hadri}},\ and\ \bibinfo {author}
  {\bibfnamefont {R.}~\bibnamefont {Khurram}},\ }\bibfield  {title} {\bibinfo
  {title} {{Scaling of a Fast Fourier Transform and a pseudo-spectral fluid
  solver up to 196608 cores}},\ }\href
  {https://doi.org/10.1016/j.jpdc.2017.10.014} {\bibfield  {journal} {\bibinfo
  {journal} {Journal of Parallel and Distributed Computing}\ }\textbf {\bibinfo
  {volume} {113}},\ \bibinfo {pages} {77} (\bibinfo {year} {2018})}\BibitemShut
  {NoStop}%
\bibitem [{Note1()}]{Note1}%
  \BibitemOpen
  \bibinfo {note} {We use Fourier band-pass filtering to determine $\protect
  \mathbf {v}_{\ell }$ and $\protect \bm {\omega }_{\ell }$ because, as is
  well-known \cite {Ott1992Sign-singularTurbulence,
  Vainshtein1994Sign-singularScalings,Sahoo2017HelicityModels}, two-point
  structure functions of vorticity are heavily influenced by small-scale
  fluctuations and thus cannot be reliably used to compare different scales in
  the inertial range.}\BibitemShut {Stop}%
\bibitem [{\citenamefont {Milanese}\ \emph {et~al.}(2020)\citenamefont
  {Milanese}, \citenamefont {Loureiro}, \citenamefont {Daschner},\ and\
  \citenamefont {Boldyrev}}]{Milanese2020DynamicTurbulence}%
  \BibitemOpen
  \bibfield  {author} {\bibinfo {author} {\bibfnamefont {L.~M.}\ \bibnamefont
  {Milanese}}, \bibinfo {author} {\bibfnamefont {N.~F.}\ \bibnamefont
  {Loureiro}}, \bibinfo {author} {\bibfnamefont {M.}~\bibnamefont {Daschner}},\
  and\ \bibinfo {author} {\bibfnamefont {S.}~\bibnamefont {Boldyrev}},\
  }\bibfield  {title} {\bibinfo {title} {{Dynamic phase alignment in inertial
  Alfv{\'{e}}n turbulence}},\ }\href
  {https://doi.org/10.1103/PhysRevLett.125.265101} {\bibfield  {journal}
  {\bibinfo  {journal} {Physical Review Letters}\ }\textbf {\bibinfo {volume}
  {125}},\ \bibinfo {pages} {265101} (\bibinfo {year} {2020})}\BibitemShut
  {NoStop}%
\bibitem [{\citenamefont {Li}(2010)}]{Li2010GeometricalTurbulences}%
  \BibitemOpen
  \bibfield  {author} {\bibinfo {author} {\bibfnamefont {Y.}~\bibnamefont
  {Li}},\ }\bibfield  {title} {\bibinfo {title} {{Geometrical statistics and
  vortex structures in helical and nonhelical turbulences}},\ }\href
  {https://doi.org/10.1063/1.3336012} {\bibfield  {journal} {\bibinfo
  {journal} {Physics of Fluids}\ }\textbf {\bibinfo {volume} {22}},\ \bibinfo
  {pages} {1} (\bibinfo {year} {2010})}\BibitemShut {NoStop}%
\bibitem [{\citenamefont {Alexakis}(2017)}]{Alexakis2017HelicallyTurbulence}%
  \BibitemOpen
  \bibfield  {author} {\bibinfo {author} {\bibfnamefont {A.}~\bibnamefont
  {Alexakis}},\ }\bibfield  {title} {\bibinfo {title} {{Helically decomposed
  turbulence}},\ }\href {https://doi.org/10.1017/jfm.2016.831} {\bibfield
  {journal} {\bibinfo  {journal} {Journal of Fluid Mechanics}\ }\textbf
  {\bibinfo {volume} {812}},\ \bibinfo {pages} {752} (\bibinfo {year}
  {2017})}\BibitemShut {NoStop}%
\bibitem [{\citenamefont {Linkmann}(2018)}]{Linkmann2018EffectsTurbulence}%
  \BibitemOpen
  \bibfield  {author} {\bibinfo {author} {\bibfnamefont {M.}~\bibnamefont
  {Linkmann}},\ }\bibfield  {title} {\bibinfo {title} {{Effects of helicity on
  dissipation in homogeneous box turbulence}},\ }\href
  {https://doi.org/10.1017/jfm.2018.709} {\bibfield  {journal} {\bibinfo
  {journal} {Journal of Fluid Mechanics}\ }\textbf {\bibinfo {volume} {856}},\
  \bibinfo {pages} {79} (\bibinfo {year} {2018})}\BibitemShut {NoStop}%
\bibitem [{\citenamefont {Buzzicotti}\ \emph {et~al.}(2018)\citenamefont
  {Buzzicotti}, \citenamefont {Aluie}, \citenamefont {Biferale},\ and\
  \citenamefont {Linkmann}}]{Buzzicotti2018EnergyRotation}%
  \BibitemOpen
  \bibfield  {author} {\bibinfo {author} {\bibfnamefont {M.}~\bibnamefont
  {Buzzicotti}}, \bibinfo {author} {\bibfnamefont {H.}~\bibnamefont {Aluie}},
  \bibinfo {author} {\bibfnamefont {L.}~\bibnamefont {Biferale}},\ and\
  \bibinfo {author} {\bibfnamefont {M.}~\bibnamefont {Linkmann}},\ }\bibfield
  {title} {\bibinfo {title} {{Energy transfer in turbulence under rotation}},\
  }\href {https://doi.org/10.1103/PhysRevFluids.3.034802} {\bibfield  {journal}
  {\bibinfo  {journal} {Physical Review Fluids}\ }\textbf {\bibinfo {volume}
  {3}},\ \bibinfo {pages} {1} (\bibinfo {year} {2018})}\BibitemShut {NoStop}%
\bibitem [{\citenamefont {Biferale}\ \emph {et~al.}(2019)\citenamefont
  {Biferale}, \citenamefont {Gustavsson},\ and\ \citenamefont
  {Scatamacchia}}]{Biferale2019HelicoidalInjection}%
  \BibitemOpen
  \bibfield  {author} {\bibinfo {author} {\bibfnamefont {L.}~\bibnamefont
  {Biferale}}, \bibinfo {author} {\bibfnamefont {K.}~\bibnamefont
  {Gustavsson}},\ and\ \bibinfo {author} {\bibfnamefont {R.}~\bibnamefont
  {Scatamacchia}},\ }\bibfield  {title} {\bibinfo {title} {{Helicoidal
  particles in turbulent flows with multi-scale helical injection}},\ }\href
  {https://doi.org/10.1017/jfm.2019.237} {\bibfield  {journal} {\bibinfo
  {journal} {Journal of Fluid Mechanics}\ }\textbf {\bibinfo {volume} {869}},\
  \bibinfo {pages} {646} (\bibinfo {year} {2019})}\BibitemShut {NoStop}%
\bibitem [{\citenamefont {Yan}\ \emph {et~al.}(2020{\natexlab{b}})\citenamefont
  {Yan}, \citenamefont {Li}, \citenamefont {Yu}, \citenamefont {Wang},\ and\
  \citenamefont {Chen}}]{Yan2020DualFlows}%
  \BibitemOpen
  \bibfield  {author} {\bibinfo {author} {\bibfnamefont {Z.}~\bibnamefont
  {Yan}}, \bibinfo {author} {\bibfnamefont {X.}~\bibnamefont {Li}}, \bibinfo
  {author} {\bibfnamefont {C.}~\bibnamefont {Yu}}, \bibinfo {author}
  {\bibfnamefont {J.}~\bibnamefont {Wang}},\ and\ \bibinfo {author}
  {\bibfnamefont {S.}~\bibnamefont {Chen}},\ }\bibfield  {title} {\bibinfo
  {title} {{Dual channels of helicity cascade in turbulent flows}},\ }\href
  {https://doi.org/10.1017/jfm.2020.289} {\bibfield  {journal} {\bibinfo
  {journal} {Journal of Fluid Mechanics}\ }\textbf {\bibinfo {volume} {894}},\
  \bibinfo {pages} {1} (\bibinfo {year} {2020}{\natexlab{b}})}\BibitemShut
  {NoStop}%
\bibitem [{\citenamefont {Waleffe}(1992)}]{Waleffe1992TheTurbulence}%
  \BibitemOpen
  \bibfield  {author} {\bibinfo {author} {\bibfnamefont {F.}~\bibnamefont
  {Waleffe}},\ }\bibfield  {title} {\bibinfo {title} {{The nature of triad
  interactions in homogeneous turbulence}},\ }\href
  {https://doi.org/10.1063/1.858309} {\bibfield  {journal} {\bibinfo  {journal}
  {Physics of Fluids A}\ }\textbf {\bibinfo {volume} {4}},\ \bibinfo {pages}
  {350} (\bibinfo {year} {1992})}\BibitemShut {NoStop}%
\bibitem [{\citenamefont {Ott}\ \emph {et~al.}(1992)\citenamefont {Ott},
  \citenamefont {Du}, \citenamefont {Sreenivasan}, \citenamefont {Juneja},\
  and\ \citenamefont {Suri}}]{Ott1992Sign-singularTurbulence}%
  \BibitemOpen
  \bibfield  {author} {\bibinfo {author} {\bibfnamefont {E.}~\bibnamefont
  {Ott}}, \bibinfo {author} {\bibfnamefont {Y.}~\bibnamefont {Du}}, \bibinfo
  {author} {\bibfnamefont {K.~R.}\ \bibnamefont {Sreenivasan}}, \bibinfo
  {author} {\bibfnamefont {A.}~\bibnamefont {Juneja}},\ and\ \bibinfo {author}
  {\bibfnamefont {A.~K.}\ \bibnamefont {Suri}},\ }\bibfield  {title} {\bibinfo
  {title} {{Sign-singular measures: Fast magnetic dynamos, and
  high-Reynolds-number fluid turbulence}},\ }\href
  {https://doi.org/10.1103/PhysRevLett.69.2654} {\bibfield  {journal} {\bibinfo
   {journal} {Physical Review Letters}\ }\textbf {\bibinfo {volume} {69}},\
  \bibinfo {pages} {2654} (\bibinfo {year} {1992})}\BibitemShut {NoStop}%
\bibitem [{\citenamefont {Vainshtein}\ \emph {et~al.}(1994)\citenamefont
  {Vainshtein}, \citenamefont {Du},\ and\ \citenamefont
  {Sreenivasan}}]{Vainshtein1994Sign-singularScalings}%
  \BibitemOpen
  \bibfield  {author} {\bibinfo {author} {\bibfnamefont {S.~I.}\ \bibnamefont
  {Vainshtein}}, \bibinfo {author} {\bibfnamefont {Y.}~\bibnamefont {Du}},\
  and\ \bibinfo {author} {\bibfnamefont {K.~R.}\ \bibnamefont {Sreenivasan}},\
  }\bibfield  {title} {\bibinfo {title} {{Sign-singular measure and its
  association with turbulent scalings}},\ }\href
  {https://doi.org/10.1103/PhysRevE.49.R2521} {\bibfield  {journal} {\bibinfo
  {journal} {Physical Review E}\ }\textbf {\bibinfo {volume} {49}},\ \bibinfo
  {pages} {R2521} (\bibinfo {year} {1994})}\BibitemShut {NoStop}%
\end{thebibliography}%


\end{document}